\documentclass{article}

\usepackage{amsthm,amsmath}
\usepackage{mathtools}
\usepackage{graphicx}
\usepackage[margin=1in]{geometry}

\RequirePackage{hyperref}

\begin{document}

\title{Predicting stock market movements using network science: An information theoretic approach}
\author{Minjun Kim$^{1,2}$ and Hiroki Sayama$^{1,2}$\\
$^1$ Department of Systems Science and Industrial Engineering\\
$^2$ Center for Collective Dynamics of Complex Systems\\
Binghamton University, State University of New York, Binghamton, NY 13902-6000\\
mkim151@binghamton.edu, sayama@binghamton.edu}

\maketitle

\begin{abstract}
A stock market is considered as one of the highly complex systems, which consists of many components whose prices move up and down without having a clear pattern. The complex nature of a stock market challenges us on making a reliable prediction of its future movements. In this paper, we aim at building a new method to forecast the future movements of Standard \& Poor's 500 Index (S\&P 500) by constructing time-series complex networks of S\&P 500 underlying companies by connecting them with links whose weights are given by the mutual information of 60-minute price movements of the pairs of the companies with the consecutive 5,340 minutes price records. We showed that the changes in the strength distributions of the networks provide an important information on the network's future movements. We built several metrics using the strength distributions and network measurements such as centrality, and we combined the best two predictors by performing a linear combination. We found that the combined predictor and the changes in S\&P 500 show a quadratic relationship, and it allows us to predict the amplitude of the one step future change in  S\&P 500. The result showed significant fluctuations in S\&P 500 Index when the combined predictor was high. In terms of making the actual index predictions, we built ARIMA models. We found that adding the network measurements into the ARIMA models improves the model accuracy. These findings are useful for financial market policy makers as an indicator based on which they can interfere with the markets before the markets make a drastic change, and for quantitative investors to improve their forecasting models.
\end{abstract}

\section{Introduction}
Stock market crashes are hard to prevent from happening due to the high complexity of the market which made of a lot of components behaving interdependently. ``The Crash of 2:45'' happened in May 6th 2010, which made U.S. Stock markets value decrease by about 6 percent in less than 30 minutes, and the flash crash occurred in Singapore took away \$6.9 billion from the Singapore Exchange are a few examples of flash crashes. Studies \cite{menk,secr,andr} including the CFTC (U.S. Commodity Futures Trading Commission) and SEC (U.S. Securities and Exchange Commission) report suggested that the main cause was said to be the high-frequency algorithmic traders dumping high volumes of the financial instruments to the market around the same time, and exacerbating the volatility during the events. Those algorithms are developed using complex mathematical models based on some theories from physics, statistics and other scientific fields with a sole purpose of producing possible trading signals. When these algorithms are triggered to make trades, the market surges or falls drastically because of the high volatility made by the algorithms in the market \cite{andr}. Considering that the high-frequency trading accounts for over 70 percent of dollar trading volumes in the U.S. financial market, and those flash crashes happened over 18,500 times between 2006 and 2011 \cite{zhan,john}, forecasting the flash crash and being able to prevent any loss are strongly needed for the unarmored ordinary individual investors' safety, healthy market ecology, and the whole economy. 

There have been many studies and developments on predicting stock market movements using many different approaches including deep learning algorithms with neural networks. Having machines learn huge sets of data such as historical stock prices, trading volumes, accounting performances, fundamental features of the stocks, and even the weather, and produce the future values of stocks or index is one big branch of stock market forecasting methods. It utilizes many learning, regression, classification, neural networks algorithms such as support vector machine, random forest, logistic regression, naive Bayes, and reccurent neural networks, and tries to make accurate predictions by adjusting itself according to the market changes \cite{erka,weih,gats,kyou}. Another popular method is to use natural language processing techniques that let machines extract and understand information written and spoken in human languages, and try to capture stock market sentiments for making investment decisions based on the mood or the sentiments of the stock market\cite{robe,hsin}. Traditional finance and modern financial engineering also attempt to forecast the stock market using the  fundamental and technical analysis. While the fundamental analysis is interested in valuating the intrinsic values of the stocks based on companies' performances and the economic status, technical analysis focuses on the price and volume dynamics, and tries to capture the investing timing by developing technical indicators \cite{wing}.

Some studies adapted network science theories to study the stock market. Those studies are mainly focused on analyzing the stock market networks’ structural properties to find out the major influencer, and to detect the communities of the stock markets \cite{anam,chik,weiq}. However, few research has been conducted to forecast the future movements of the stock market using networks science. One of the few studies built corporate news networks using top 50 European companies in STOXX 50 index as nodes, and the sum of the number of news items with the common topic of each company pair as link weights. This study found out that the average eigenvector centrality of the news networks has an impact on return and volatility of the STOXX 50 index \cite{germ}. Another study constructed a role-based trading network for each company characterizing the daily trading relationship among its investors with transaction data. Particularly, nodes are traders involved in the transactions of a stock, and for each transaction between two traders, there is a link from the seller to the buyer. By categorizing the nodes into three types (Hub, Periphery, and Connector) according to the node's connectedness, this study created 9 different link types, and found that the time-series of fraction of the link type P-H and C-H have a predictive power with the maximum accuracy of 69.2\% \cite{xiao}.

Network science has been used and developed for many different fields. However, a few studies were conducted in terms of financial market time-series forecasting. Also, the previous studies did not show that whether the network analysis helps improve the performances of financial market time-series forecasting models. In this paper, we discuss our network analysis that forecasts future amplitudes of the S\&P 500 changes to the one hour future and helps improve the performances of ARIMA models.

\section{Method}
This study aims at building a reliable stock market prediction model based on the stock market networks analysis, and prove that the network measurements indeed improve the ARIMA models. The study follows the following steps. First,  we acquire the raw data set of stock price records. Second, we pre-process the data set with an appropriate imputation method to deal with missing data. Third, we compute the pairwise mutual information of the stocks in S\&P 500. Fourth,  we construct networks using the mutual information as link weights. Fifth, we compute the node strength distribution of each network. Sixth, we build the several predictors using the strength distribution, network centrality and modularity. Seventh, we build linearly combined predictors with the two best performing metrics that maximizes the correlation between the metric and the changes in S\&P 500. Eighth, we built ARIMA models to predict the actual S\&P 500 index. Lastly, we show whether or not the network measurements help improve the ARIMA models. More details are discussed in the following subsections.

\subsection{Data gathering and preprocessing}
Among the 504 companies in S\&P 500 components at the time of the analysis, 475 companies' stock price records were used due to the data availability. Also, S\&P 500 index record was used as the dependent variable of our model. Each time-series record consists of 5,340 one-minute interval closed prices ranging from 9:30am in September 22nd 2016 to 4:00pm in October 11th 2016, which is for 89 consecutive trading hours. The stock price records for 475 stocks and S\&P 500 index records were acquired from the Google Finance (https://www.google.com/finance) real-time price quotes. The data set contained missing values due to the fact that some stocks were not traded at specific dates or times where the stock exchanges halted or delayed the trades of the specific companies’ stocks for news pending or significant imbalances in the pending buy and sell orders \cite{chri}. The data set was preprocessed to handle the missing data in hot deck imputation method - ``last observation carried forward'' - particularly by replacing the missing values with the nearest available data points in the same time-series specifically with the very last closed price \cite{dice}. 

\subsection{Network construction}
With the time-series data for each company in the S\&P 500, we formed hourly networks in order to inspect the hourly movements of S\&P 500 and make a prediction to the one step forward. In total, 89 hourly networks were constructed with 475 nodes representing companies of S\&P 500 and edges representing the pairwise mutual information of 60-minute price movements of the stocks. Some previous studies used the correlation as link weights when constructing networks \cite{anam,weiq}, but because the mutual information is better at capturing the non-linear correlation \cite{liwe}, we chose to use the mutual information as our network link weights. The mutual information measures the amount of information of a variable given the information of another variable, and tells us the non-linear correlation of the two variables. Particularly for our analysis, because we investigated one-hour movement of the S\&P 500 with one-minute interval data, we split the 5,340-minute long time-series data into 60-minute long non-overlapping windows for each stock. For each window, we computed mutual information of every stock pair by the equation \eqref{equation1} with a binning operation (bin size = 5) where $X_i$ and $Y_j$ are price records of a pair stocks of the specific window, and $p(x)$ is the probability of a random sample $x$ occurring in $X_i$, and $p(y)$ is the probability of a random sample $y$ occurring in $Y_j$, and $ i = \left\{s_1, s_2, \cdots, s_{475}\right\} $ and $ j = \left\{s_1, s_2, \cdots, s_{475}\right\}$ representing 475 companies, while $p(x,y)$ is the joint probability. Finally, we created mutual information matrix for all the stock pairs as shown below matrix \eqref{matrix1}.

\begin{align}
I(X_i;Y_j )&= \sum_{x \in X_i, y \in Y_j, x \ne y} p(x,y) log\left(\frac{p(x,y)}{p(x) p(y)}\right), \label{equation1}
\end{align}

\begin{align}
M &=\displaystyle\bordermatrix{~ &  Y_{s1} & Y_{s2} & \cdots & Y_{s475} \cr
X_{s1}&  0 & I(X_{s1};Y_{s2}) & \cdots & I(X_{s1};Y_{s475}) \cr
X_{s2}&  I(X_{s2};Y_{s1}) & 0 & \cdots & I(X_{s2};Y_{s475}) \cr
 \hfil\hfil\vdots\hfil & \vdots & \vdots & \ddots & \vdots \cr
X_{s475}&  I(X_{s475};Y_{s1}) & I(X_{s475};Y_{s2}) & \cdots & 0 \cr}. \label{matrix1}
\end{align}

\smallskip
As for the network links, we assigned a link weight with the corresponding mutual information. This means that the networks are complete weighted graphs having links between every node and every other node, and since the mutual information is symmetric, the network is also undirected. Some previous studies formed the networks with a threshold value on the link weights for a specific purpose of detecting the stock market clusters to see whether the stocks in the same industries or sectors fall into the same community as of their market classifications \cite{anam}. However, in this particular study, we took all the link weights under our consideration in order to study the dynamics of the strength distributions of the networks for forecasting the future stock market fluctuations.

\subsection{Analysis}
\begin{figure}[tb]
\centering
\includegraphics[width=6in]{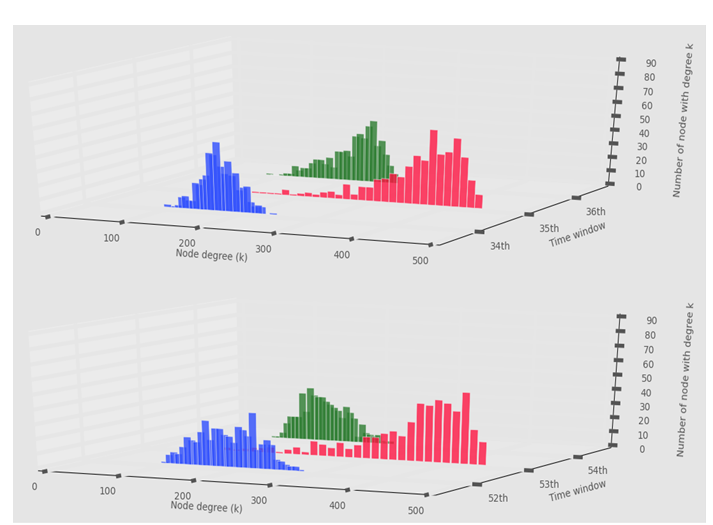}
      \caption{\textbf{Unusual deviations in Strength distribution.}
      The strength distribution(blue) drastically shifts to the right(pink) having more high strength nodes, and comes back to the normal range(green). Because these deviations are followed by large changes in S\&P 500, they could be used as a representation of the dynamics of S\&P 500 index.}
       \label{fig:degreedistfortwo}
         \end{figure}

We formed several different metrics that will be used as our predictors to forecast the amplitude of the future S\&P 500 changes using the strength distributions and the network measurements such as the network centrality and modularity. We first computed the strength of each node for each network, and formed the strength distributions for all the 89 networks. The strength distributions were normally on the node strength range of $0 \sim 300$, but sometimes it went over 300 with different distribution shapes and more high strength nodes. For example, the Figure~\ref{fig:degreedistfortwo} shows unusual movements of the strength distributions shifting to the right and coming back to the normal range. These movements are critical information for our study because they are followed by large changes in S\&P 500 index. In order to investigate this founding thoroughly, we formed several metrics using the strength distributions. First, we formed a metric by interpreting the distribution shapes into numeric values using Kullback-Leibler divergence (KLD). Second, we formed the relative strength (RS) metric using the actual node strength data. Lastly, we formed a couple of statistical  metrics such as mean, variance, skewness and kurtosis of the strength distributions. In addition, we computed other network measurements to see the relationship with S\&P 500 index. More details are described in the following subsections. All statistical measures for the analyses in the results section are computed using statistics tools (networkX, scikit-learn and statmodels) in Python.

\begin{figure}[tbp]
\centering
\includegraphics[width=6.5in]{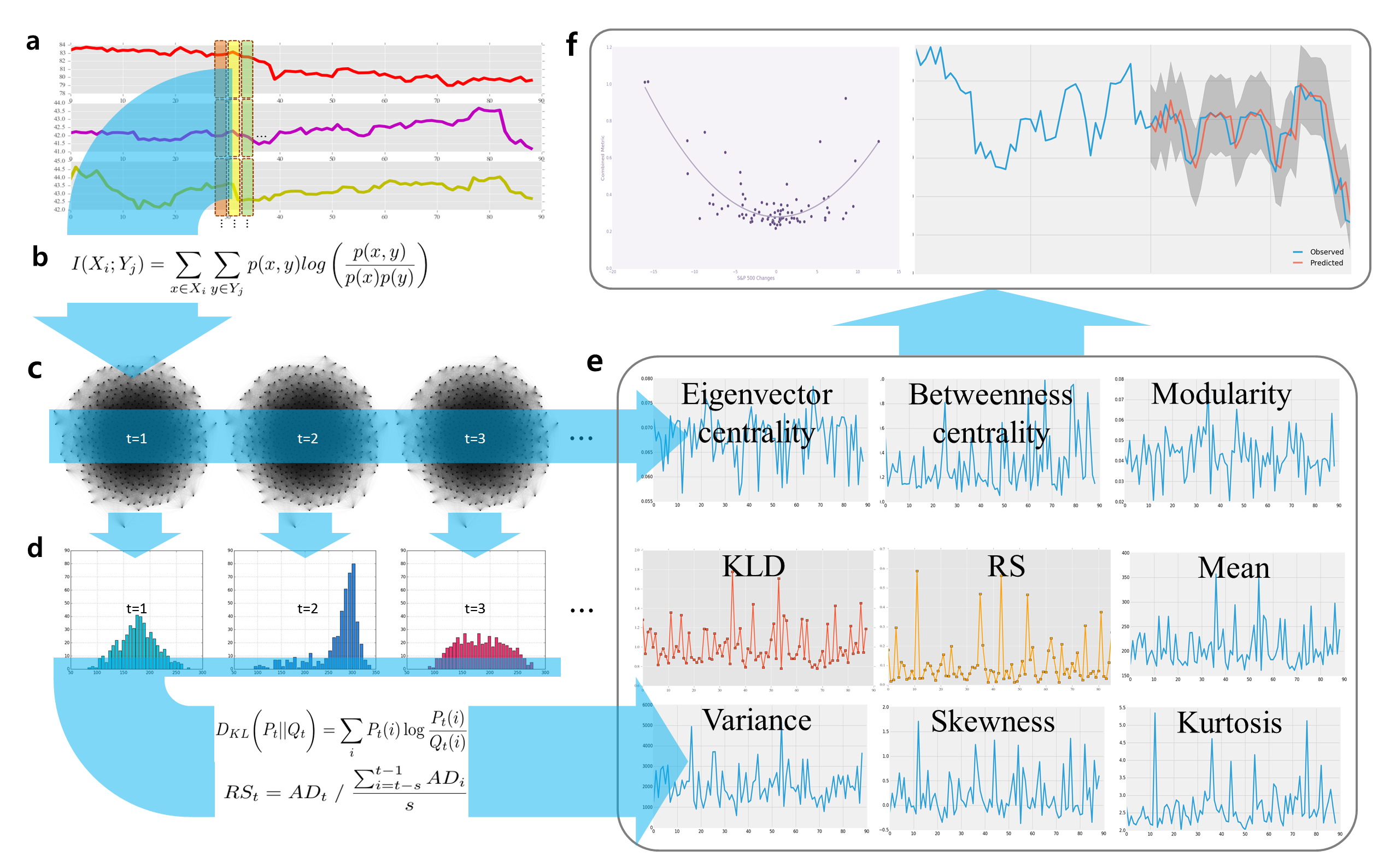}
      \caption{\textbf{Analysis Process}
      \textbf{a}. gather time-series stock price records of the S\&P 500 underlying companies, and split the records by one hour records,  \textbf{b}. calculate mutual information of the stock pairs,  \textbf{c}. construct networks using the mutual information as link weights,  \textbf{d}. computing the strength distribution for each network,  \textbf{e}. build metrics with the strength distribution data and other network measurements (average, median and maximum values of eigenvector and betweenness centralities are used). The plotted centralities are maximum values of centralities,  \textbf{f}. predict the amplitude of S\&P 500 changes by forming a linear combination of top performing metrics, and forecast actual S\&P 500 index by building ARIMA models with network measurements.}
       \label{fig:kldandfreqncy}
         \end{figure}

\subsubsection{Metrics using strength distributions}

The first metric was formed by using Kullback-Leibler divergence which is to compute the relative entropy of each network against the average prior distribution $(Q_t)$, the probability distribution of the aggregated average strength of the prior networks calculated by the Eq. \eqref{eq1} where $s$ is the number of prior networks. Kullback-Leibler divergence, which is also called relative entropy, is a measure of the difference between two probability distributions $P$ and $Q$ where $P$ is the distribution of the observation that we want to see how much it differs from the average prior distribution $Q$. For example, $Q_t$ for calculating KLD-$s$ of the t $th$ network is the average of the $t$-1 $th$, $t$-2 $th$, $\cdots$, and  $t$-s $th$ networks' strength distributions. Considering that the stock market runs 6.5 hours daily, we computed Kullback-Leibler divergence between each strength distribution and the average of 3 hours (0.5-trading day), 6 hours (1-trading day), 9 hours (1.5-trading day), 13 hours (2-trading days) and all period prior distributions namely KLD-3, KLD-6, KLD-9, KLD-13 and KLD-All. The Kullback-Leibler divergence of $P$ from $Q$ was calculated by Eq. \eqref{eq2}.

\begin{align}
Q_t &= \frac{\sum_{i=t-s}^{t-1} P_i}{s},\label{eq1} \\
D_{KL}\biggl( P_{t}||Q_t\biggr) &= \sum_i P_{t}(i)\log \frac{P_{t}(i)}{Q_t(i)}.\label{eq2}
\end{align}

\smallskip
The second metric was formed by using the actual node strength instead of using probability distributions. We call this metric relative strength ($RS$) of the network. We calculated the average strength ($AD$) of nodes of each network, and divided it by the average strength of prior networks as shown in the Eq. \eqref{eq5} where $s$ is the number of prior networks.

\begin{align}
RS_t &= {AD_t}\hspace{0.1cm}/\hspace{0.1cm} {\frac{\sum_{i=t-s}^{t-1} AD_i}{s}}.\label{eq5}
\end{align}

\smallskip
\subsubsection{Network centrality and modularity}

Network centrality is often used in social network analysis for finding out the most influential or important nodes in networks by measuring their cohesiveness or involvements in the networks. In the sense that our stock market networks contained more high strength nodes when there were large changes in S\&P 500 index, metrics using centrality of the nodes could explain the movements of S\&P 500 index. We computed eigenvector and betweenness centralities of the nodes in our S\&P 500 networks, and formed metrics using their mean, median and maximum values. In the same sense, we also formed metrics using network modularity. Modularity is a clustering measure that is to find the community structure of a network. A high modularity means that there are more links within a specific group in a network than when links are randomly distributed among groups in the network. 

For each metric we built, we tested their prediction performance against the actual changes, squares of the changes and absolute values of the changes in S\&P 500 index. The second and third-order of polynomial regression and a simple linear regressions are used for performing the fitting tests. Results are shown in the results section.

\section{Results}

We tested 21 different metrics for forecasting the amplitudes of S\&P 500 changes using network measurements. First of all, in all cases, there was no strong linear relationship between the metrics and the actual changes in S\&P 500. This was because that, in most cases, the metrics and S\&P 500 changes had a quadratic relationship. This was followed by the fact that the nodes were more strongly connected to each other both when S\&P 500 went up or down. This fact is shown with the higher correlations in Act. S\&P * and Abs. S\&P ** cases (see Table~\ref{table:correlation}). The correlation between the metrics and S\&P 500 changes were much stronger in the polynomial regressions with actual ups and downs of the index, and linear regression with absolutes changes of the index. 

\begin{table}[tbp]
\centering
\caption{Correlation coefficients between S\&P 500 changes and the predictors with polynomial and linear regressions.}
     \begin{tabular}{ccccc}
        \hline
        Correlation matrix  & Act. S\&P * & Act. S\&P ** & Sqrs. S\&P ** &  Abs. S\&P ** \\ \hline
         & & & & \\
        Strength distribution & & & & \\\hline
         & & & & \\
         
        KLD 3 & 0.5628 / 0.6081 & 0.0895 & 0.6705 & 0.6454 \\ 
        KLD 6 & 0.5752 / 0.5984 & 0.0837 & 0.6823 & 0.6360 \\ 
        KLD 9 & 0.5333 / 0.5334 & 0.0582 & 0.6498 & 0.6725 \\ 
        KLD 13 & 0.4794 / 0.4886& 0.0408 & 0.6182 & 0.6219 \\ 
        KLD All & 0.5582 / 0.5635 & 0.1175 & 0.6521 & 0.6587 \\ 
         & & & & \\ 
         
	RS 3 & 0.4185 / 0.4455 & 0.0173 & 0.3811 & 0.6630 \\ 
        RS 6 & 0.4326 / 0.4845 & 0.0159 & 0.3838 & 0.6615 \\ 
        RS 9 & 0.4196 / 0.4855 & 0.0093 & 0.3750 & 0.6526 \\ 
        RS 13 & 0.4385 / 0.4674 & 0.0024 & 0.4077 & 0.6685 \\ 
        RS All & 0.4065 / 0.4447 & 0.0134 & 0.3649 & 0.6552 \\
         & & & & \\
         
        Mean & 0.4189 / 0.4583 & 0.0129 & 0.3640 & 0.6536 \\ 
        Variance & 0.1487 / 0.1641 & 0.0175 & 0.3548 & 0.6407 \\ 
        Skewness & 0.5471 / 0.5610 & 0.0644 & 0.6265 & 0.5716 \\ 
        Kurtosis & 0.5425 / 0.5581 & 0.0192 & 0.4047 & 0.6532 \\
         & & & & \\
        
        Eigenvector centrality & & & & \\\hline
        Mean & 0.1526 / 0.1795 & 0.0099 & 0.2591 & 0.5351 \\ 
        Median & 0.3168 / 0.3200 & 0.0110 & 0.2720 & 0.5509 \\ 
        Maximum & 0.4272 / 0.4494 & 0.0068 & 0.2175 & 0.4836 \\ 
         & & & & \\
         
        Betweenness centrality & & & & \\\hline 
        Mean & 0.0435 / 0.0482 & 0.0111 & 0.2797 & 0.5482 \\ 
        Median & 0.0288 / 0.0289& 0.0102 & 0.0089 & 0.0332 \\
        Maximum & 0.0288 / 0.0288 & 0.0162 & 0.2445 & 0.4350\\
         & & & & \\
         
        Network modularity & & & & \\\hline 
        Modularity & 0.2973 / 0.2982 & 0.0082 & 0.1503 & 0.3906 \\ \hline
      \end{tabular} \\
      \vspace{1 ex}
      \hspace{0.4cm}\textbf{Note.}            
      \raggedright Correlations are significant at the 0.05 level.
      
      \hspace{1.1cm} Act.: Actual values of S\&P 500 changes.
      
      \hspace{1.1cm} Sqrs.: Squared values of S\&P 500 changes.
      
      \hspace{1.1cm} Abs.: Absolute values of S\&P 500 changes.
      
      \hspace{1.1cm} * Polynomial regression second-order/third-order.
      
      \hspace{1.1cm} ** Linear regression.
      \label{table:correlation}
\end{table}

\begin{figure}[tbp]
\centering
\includegraphics[width=6.5in]{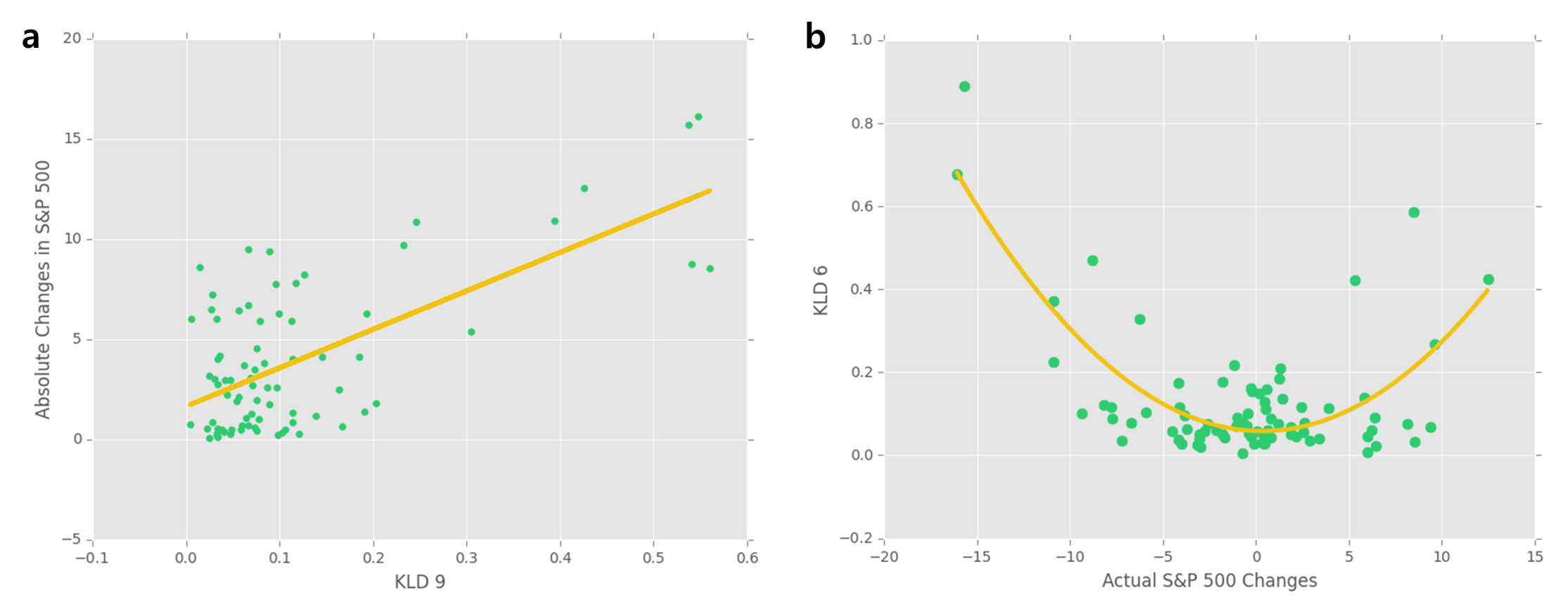}
      \caption{\textbf{Kullback-Leibler divergence and S\&P 500 Changes}
       \textbf{a}. KLD 9 has the strongest linear correlation (R-squared = 0.6725) with absolute values of S\&P 500 changes among all KLD metrics.  \textbf{b}. KLD 6 has the strongest quadratic correlation with the actual S\&P 500 Changes (R-squared = 0.5752). These can be used to forecast the amplitude of the S\&P 500 changes.}
       \label{fig:kldscatterwsp500}
         \end{figure}

The KLD metrics were performing better than other metrics in most of the cases, having the maximum r-squared of 0.6725 for Abs. S\&P ** case with KLD 9. Figure~\ref{fig:kldandfreqncy} is a graph of all the KLD metrics with actual and absolute values of S\&P 500, showing that the divergences were normally below the 0.2 and there were some hikes from time to time. A low $KLD$ means the nodes strength of a network at a specific time were distributed similar to the average prior distribution, and a hike means that the strength distribution of the network drastically deviates from the average prior distribution, having more high strength nodes. This reflects that when $KLD$ hikes, the movements of the stocks in the networks have more correlation in each other's movements. As we can see in this graph and the scatter plot(see Figure~\ref{fig:kldscatterwsp500}) KLD has a clear positive correlation with absolute values of S\&P 500 changes, and shows a quadratic relationship having both negative and positive relationships with the actual changes in S\&P 500.

\begin{figure}[tbp]
\centering
\includegraphics[width=4.5in]{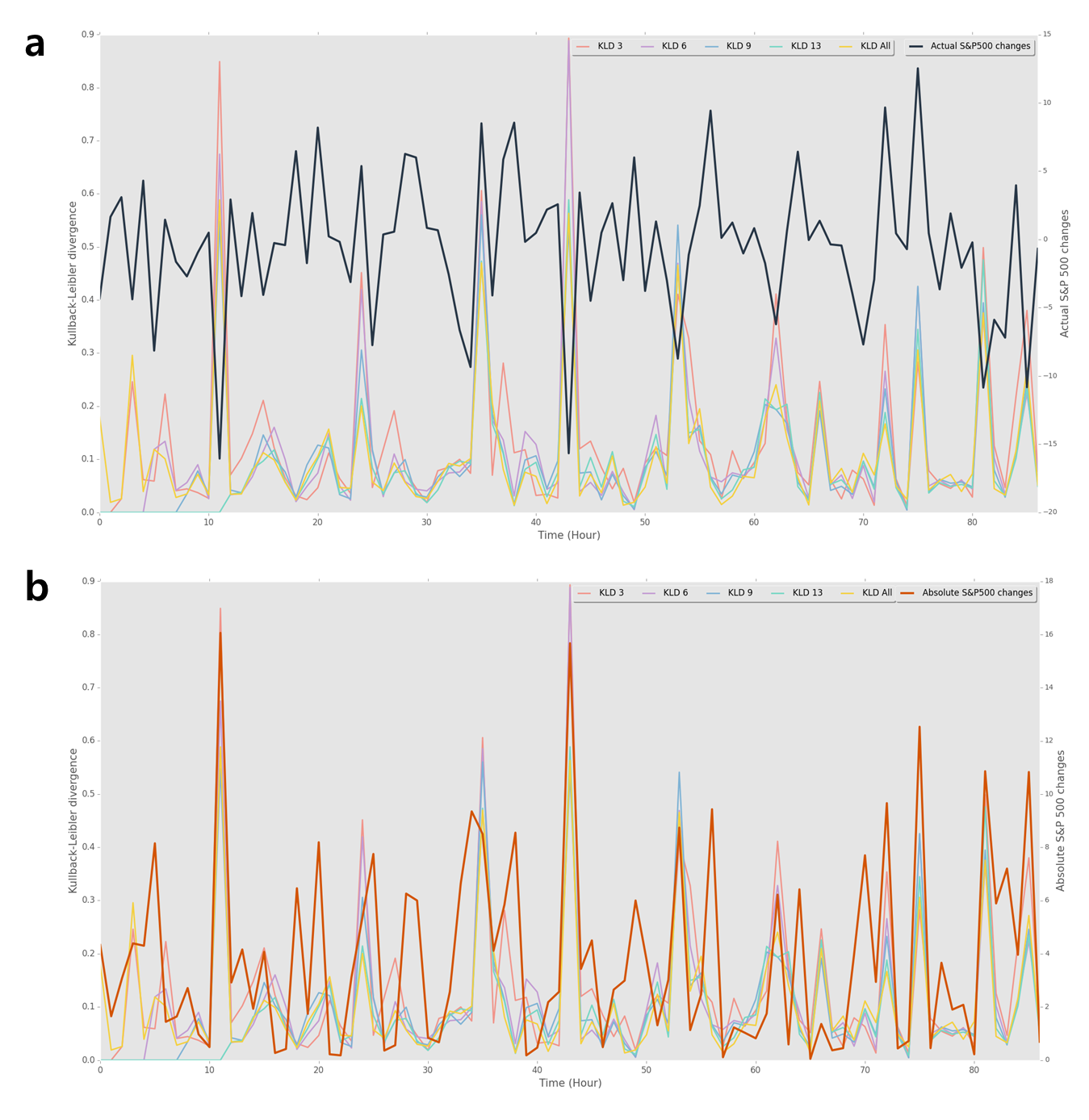}
      \caption{\textbf{Kullback-Leibler divergence and S\&P 500 changes}
       \textbf{a}. KLD normally stays between 0.0 and 0.2, but sometimes shows significant hikes. KLD and actual S\&P 500 changes are negatively and positively correlated.  \textbf{b}. KLD and absolute values of S\&P 500 changes show a clear positive correlation. It is harder to predict the actual changes than to predict amplitude of the changes.}
       \label{fig:kldandfreqncy}
         \end{figure}

\begin{figure}[tbp]
\centering
\includegraphics[width=4.5in]{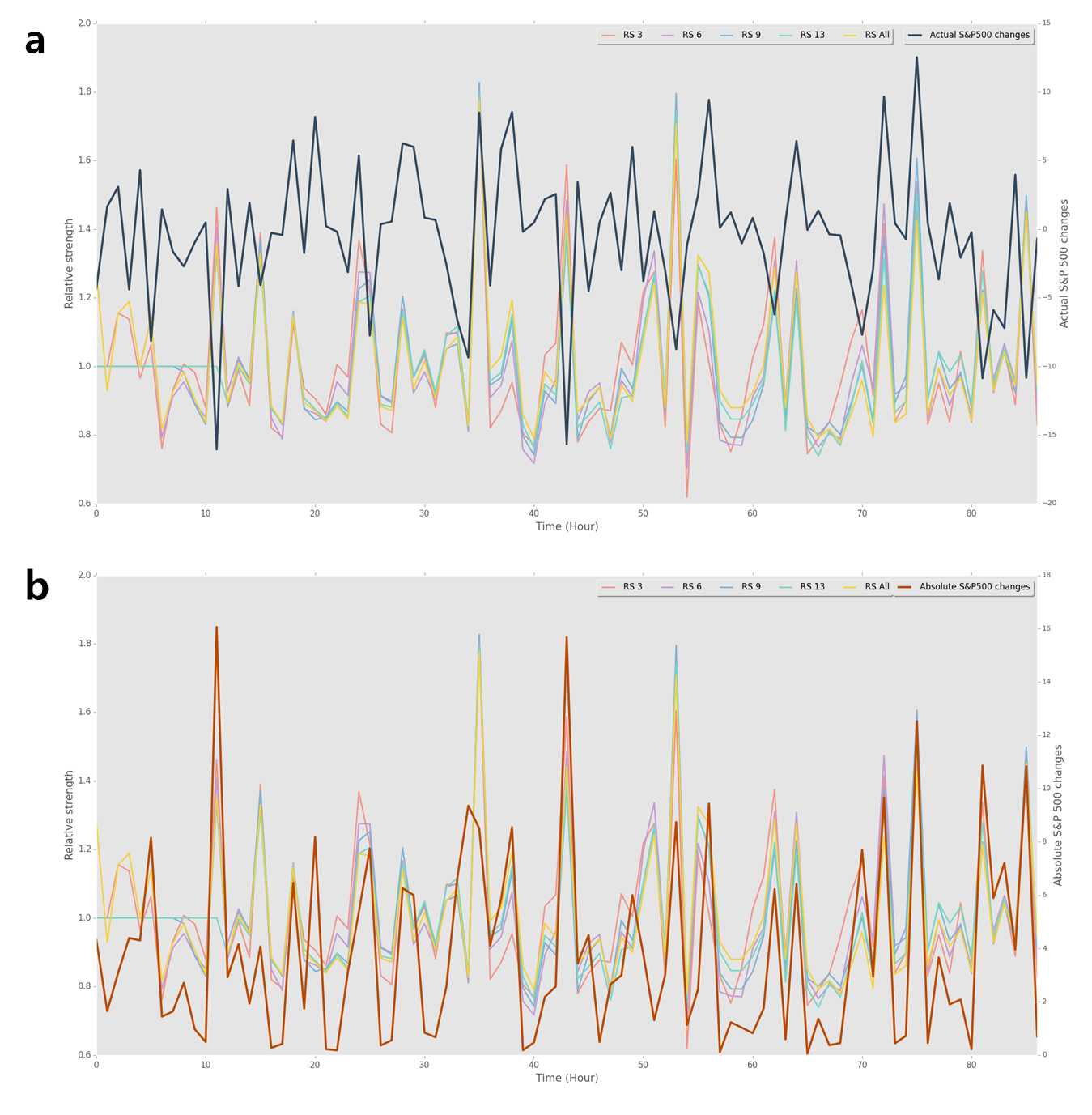}
      \caption{\textbf{Relative strength and S\&P 500 changes}
       \textbf{a}. RS normally stays between 0.8 and 1.0, but sometimes shows significant hikes. RS and actual S\&P 500 changes are negatively and positively correlated.  \textbf{b}. RS and absolute values of S\&P 500 changes show a clear positive correlation.}
       \label{fig:rsline}
         \end{figure}

\begin{figure}[tbp]
\centering
\includegraphics[width=6.5in]{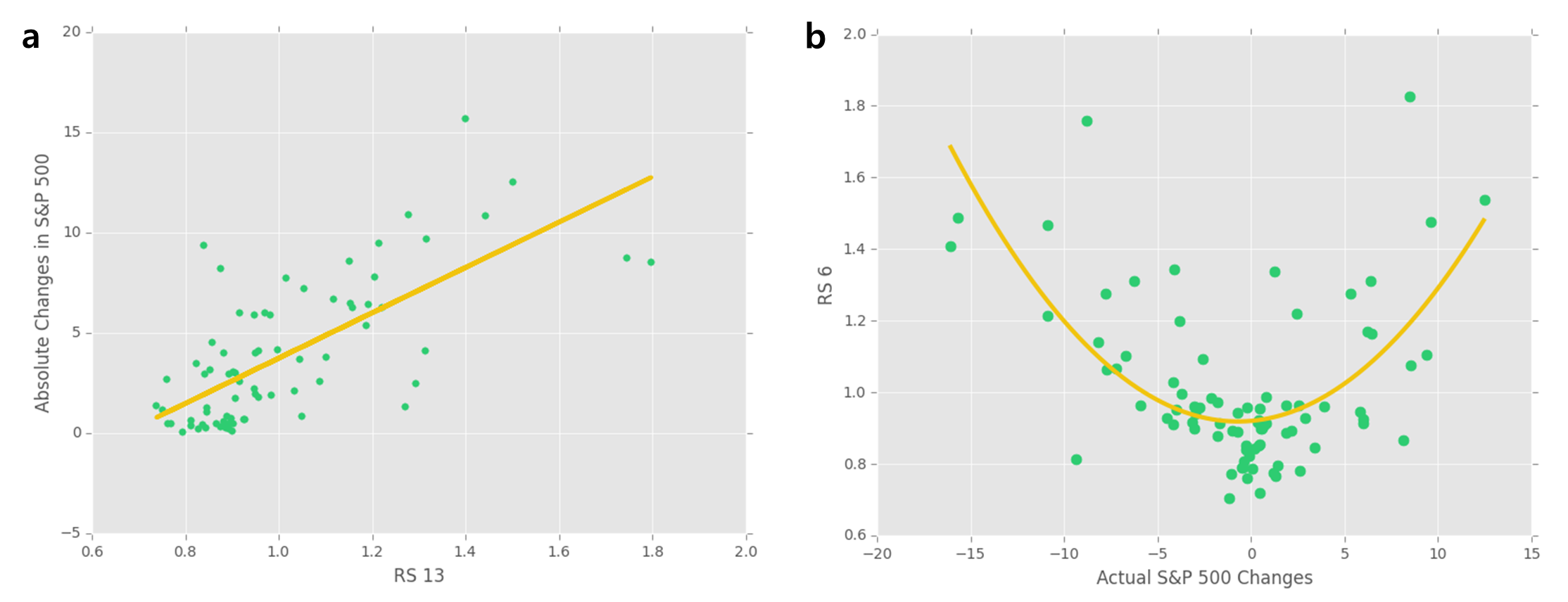}
      \caption{\textbf{Relative strength and S\&P 500 Changes}
       \textbf{a}. RS 13 has the strongest linear correlation (R-squared = 0.6685) with absolute values of S\&P 500 changes among all RS metrics.  \textbf{b}. RS 6 has the strongest quadratic correlation with the actual S\&P 500 Changes (R-squared = 0.4326).}
       \label{fig:rsscat}
         \end{figure}

Similar to the Kullback-Leibler divergence, Relative Strength also showed a quadratic relationship with actual S\&P 500 changes and a clear linear relationship with absolute values of the changes (see Figure~\ref{fig:rsline} and Figure~\ref{fig:rsscat}). Relative strength is normally in between $0.8$ and $1.1$ which indicates that the strength of the networks are usually similar to that of the prior distributions. However, there were number of cases where the relative strength was much higher or lower from time to time. The overall prediction performance was worse than KLD's, but for predicting the absolute values of the changes with linear regressions, it performed as good as KLD.

Among the statistical measures using the strength distributions, the skewness was the top performer in Act. S\&P 500 * case (R-squared = 0.5471), and the kurtosis was the top performer in Abs. S\&P 500 ** case (R-squared = 0.6532). The mean and variance performed much better in predicting the absolute changes in S\&P 500 than predicting the actual changes. 

We tested average, median, and maximum values of eigenvector and betweeness centralities. They showed correlations with actual and absolute values of the S\&P 500 changes. Especially, predicting the absolute values of the changes in S\&P 500, it showed R-squared of 0.5. This relationship was not strong, but still explained that dynamics of the S\&P 500 index can be explained by the structural property of the stock market networks - when the nodes in the networks were clustered, grouped or tied more strongly, there came the large changes in S\&P 500 index. Modularity, however, performed poorer than other metrics having no clear relationships with the S\&P 500 changes. 

To make stronger predictors out of all the metrics, we constructed linear combinations. We used two different ways for predicting the amplitude. First, we used a polynomial regression on the actual values of the changes. Second, we used a linear regression on the absolute values of the changes. We picked the top two performers for each method to perform the linear combination. We picked KLD 6 and Skewness(S), and KLD 3 and Skewness for the polynomial regression degree 2 and 3 for the first method, and picked KLD 9 and RS 13 for the second method. Particularly, we form three  combined metrics $C1 = a  KLD3 + (1-a)  S$, $C2 = a  KLD6 + (1-a)  S$, and $C3 = a  KLD9 + (1-a)  RS13$. We optimized the  two constants to have a maximum correlation between the combined metrics and the S\&P 500 changes by performing the grid search method for finding out the optimal value of $a$. We found that all of the three metrics have statistically significant correlations with the S\&P 500 changes. Table~\ref{table:combinedpred} shows the correlations between S\&P 500 changes and the three predictors as well as the optimal value of the constant $a$. $C3$ performed the best in predicting the amplitudes of the S\&P 500 changes. It showed R-squared of 0.7301 in the linear regression with the optimized constant $a=0.834$. $C2$ and $C1$ in the polynomial regressions explained about 64\% of the variance in the actual values of S\&P 500 changes. As seen in this result, we could predict the amplitude better when working with the absolute values of S\&P 500 changes with linear regression rather than working with the actual values with polynomial regressions. 

\begin{table}[tbp]
\centering
\caption{Combined predictors vs. S\&P 500 changes.}
     \begin{tabular}{ccccc}
        \hline
           Combinations & Constant($a$) & Linear & Polynomial (k=2) & Polynomial (k=3) \\ \hline  
        KLD3 + Skewness & 0.798 & - & - & 0.6481 \\
        KLD6 + Skewness & 0.767 & - & 0.6409 & - \\
        KLD9 + RS13 & 0.834 & 0.73012 & - & -  \\ \hline
      \end{tabular} \\
      \vspace{1ex}
      \hspace{0.65cm}\textbf{Note.}                  
      \raggedright Correlations are significant at the 0.05 level.

      \label{table:combinedpred}
\end{table}

\begin{figure}[tbp]
\centering
\includegraphics[width=4.5in]{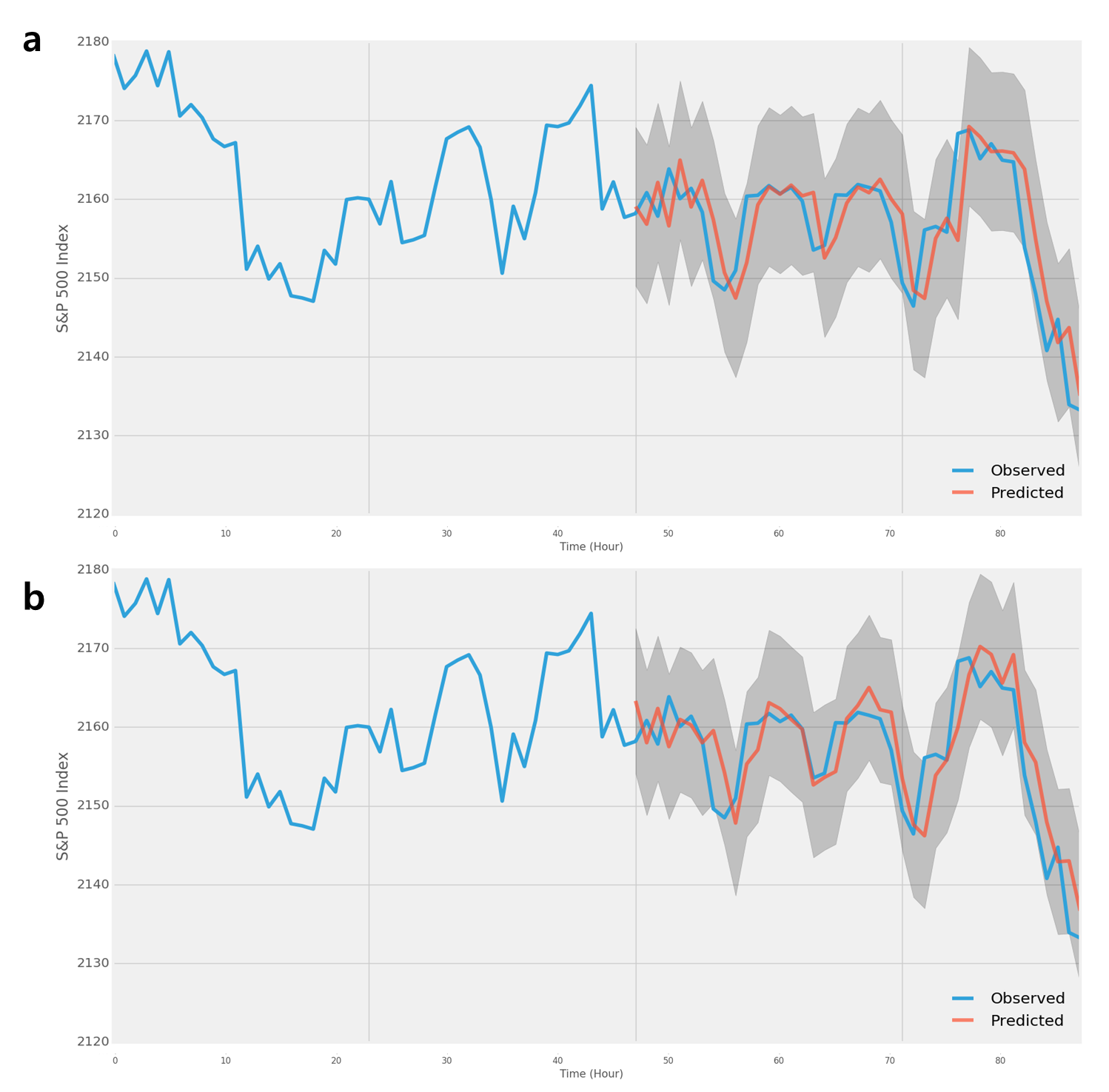}
      \caption{\textbf{ARIMA models actual vs. predicted S\&P 500 index.}
      \textbf{a}. $ARIMA(1,1,1)$ looks like a 1-lag moving average of the actual S\&P 500 series. This model performed MSE of 25.19. \textbf{b}. $ARIMA(1,1,0)$ with RS model lowered the gap between the actual and predicted series, and it performed MSE of 20.27 which was lower than $ARIMA(1,1,1)$ model by 20\%.}
       \label{fig:ARIMA111}
         \end{figure}

In order to investigate the prediction power of the network measurements on predicting the actual S\&P 500 index, we built several autoregressive integrated moving average models (ARIMA). ARIMA is one of the popular time-series forecasting models in statistics, and often used for financial market time-series forecasting. We first built an ARIMA model which only took an endogenous variable ,historical S\&P 500 index that we used earlier. We performed a grid search for optimizing the three ARIMA parameters (p, d, q): p for the order of autoregressive model, d for the degree of differencing, and q for the order of the moving average model. Out of all ARIMA models, we chose ARIMA(1,1,1) for the further analysis because it yielded the lowest Akaike information criterion (AIC) which is a measure of a statistical model considering the goodness of fit and simplicity of the model. Figure~\ref{fig:ARIMA111}-a shows the actual and predicted values of S\&P 500 index by the ARIMA(1,1,1) model. The model looked pretty much like a 1-lag moving average of the original series, and performed a mean squared error of 25.19. To see if the network measurements are useful to be added in ARIMA models, we included each and every network measurements into the ARIMA models as an exogenous variable. Table~\ref{table:ARIMA_MSE} shows the mean squared error of the ARIMA models with and without network measurements. Based on the results, ARIMA(1,1,0) with RS model lowered the MSE by 4.92 which was a significant improvement. ARIMA(1,1,0) with betweenness centrality also lowered the MSE. As we can see from the Figure~\ref{fig:ARIMA111}-b, the predictions from ARIMA(1,1,0) with RS got closer to the actual values of S\&P 500 index. 

\begin{table}[tbp]
\centering
\caption{Mean Squared Errors of ARIMA models.}
     \begin{tabular}{cccc}
        \hline
          Models & MSE*  & Models  & MSE* \\ \hline
        ARIMA(1,1,1) & 25.19 & ARIMA(1,1,0) + KLD & 26.17\\
        ARIMA(1,1,0) + RS & 20.27 & ARIMA(1,1,0) + Skewness & 26.32 \\
        ARIMA(1,1,0) + Kurtosis  & 27.02 & ARIMA(1,1,0) + Mean & 25.7 \\
        ARIMA(1,1,0) + Variance & 25.66 & ARIMA(1,1,1) + Modularity & 25.23 \\
        ARIMA(1,1,0) + Eigenvector cent & 26.95 & ARIMA(1,1,0) + Betweenness cent & 24.4 \\ \hline
      \end{tabular} \\
      \vspace{1ex}
      \hspace{0.45cm}\textbf{Note.}                  
      \raggedright * Mean Squared Error
      \label{table:ARIMA_MSE}
\end{table}

To sum up, some of the network measurements we built in this research have forecasting power on predicting the amplitudes of S\&P 500 changes. KLD, RS and Skewness of the strength distributions were the top performers with the significant correlations of over 0.64. Also, adding RS into the ARIMA model improved the model performance by about 20\%. 

\nocite{anam,secr,andr,wing,germ,xiao,robe,hsin,chik,weiq,erka,weih,cliv,gats,kyou,newm,chri,dice,liwe,saya,bara,vinh,havr,easl,menk,zhan,john}

\section{Discussions and Conclusions}
In this study, we demonstrated a new approach to forecast future S\&P 500 changes using networks science, and showed that the predictors we built were strongly correlated to the amplitude of the S\&P 500 changes. This result was because that we could be able to capture the market dynamics by analyzing the S\&P 500 networks. The networks showing high connectedness among all the companies(nodes) means the stocks are more highly correlated. Stocks are highly correlated when the stocks are bought or sold together. As a whole market point of view, when most of the stocks are moving together, the index is likely to move in the same direction as the majority of stocks is moving. Our ARIMA models were improved by adding RS. The results can be used as a new indicator that might advise financial policy makers in dealing with huge sudden market fluctuations that definitely bring the market serious problems. Also, the result can be used for the quantitative investors to improve their existing ARIMA models. In this paper, we tested the usefulness of network measurements with ARIMA model only. However, in the future, we will investigate whether the network measurements help improve other financial market time-series forecasting models such as machine learning models.

In this study, we were able to get 475 companies' stock price records out of 504 companies. It might be possible to improve the performance of the models if we have the price records for all the companies. Another improvement can be achieved by using finer data such as a half minute interval price records or even finer than a half minute. If we use finer data sets, we might be able to improve the model for forecasting one hour future, and also able to forecast nearer future such as 30-minute future or 15-minute future.
 
\bibliographystyle{plain}
\bibliography{kim-sayama}

\end{document}